\documentclass[useAMS,usenatbib]{mn2e}
\usepackage{graphicx,amssymb}

\newcommand{\bc}{\begin{center}}
\newcommand{\ec}{\end{center}}

\title[Chemical enrichment in a $\Lambda$CDM model]
      {Chemical enrichment of the intra--cluster and intergalactic medium in
        a hierarchical galaxy formation model}
\author[G.~De Lucia et al.]
        {Gabriella De Lucia\thanks{Email: gdelucia@mpa-garching.mpg.de},
        Guinevere Kauffmann and 
        Simon D.~M.~White     
        \\      
        Max--Planck--Institut f\"ur Astrophysik, 
        Karl--Schwarzschild--Str. 1, D-85748 Garching, Germany}
\begin{document}

\date{Accepted 2003 ???? ??. 
      Received 2003 ???? ??; 
      in original form 2003 April 28}

\pagerange{\pageref{firstpage}--\pageref{lastpage}} 
\pubyear{2003}

\maketitle

\label{firstpage}

\begin{abstract}
  We use a combination of high resolution $N$--body simulations and
  semi--analytic techniques to follow the formation, the evolution and the
  chemical enrichment of galaxies in a $\Lambda$CDM Universe.  We model the
  transport of metals between the stars, the cold gas in galaxies, the hot gas
  in dark matter haloes, and the intergalactic gas outside virialized haloes.
  We have compared three different feedback schemes. The `retention' model
  assumes that material reheated by supernova explosions is able to leave the
  galaxy, but not the dark matter halo.  The `ejection' model assumes that this
  material leaves the halo and is then re--incorporated when structure
  collapses on larger scales. The `wind' model uses prescriptions that are
  motivated by observations of local starburst galaxies.  We require that our
  models reproduce the cluster galaxy luminosity function measured from the
  $2$dF survey, the relations between stellar mass, gas mass and metallicity
  inferred from new SDSS data, and the observed amount of metals in the ICM.
  With suitable adjustment of the free parameters in the model, a reasonable
  fit to the observational results at redshift zero can be obtained for all
  three feedback schemes.  All three predict that the chemical enrichment of
  the ICM occurs at high redshift: $60$--$80$ per cent of the metals currently
  in the ICM were ejected at redshifts larger than $1$, $35$--$60$ per cent at
  redshifts larger than $2$ and $20$--$45$ per cent at redshifts larger than
  $3$.  Massive galaxies are important contributors to the chemical pollution:
  about half of the metals today present in the ICM were ejected by galaxies
  with baryonic masses larger than $10^{10}\,h^{-1}\,{\rm M}_{\odot}$.  The
  observed decline in baryon fraction from rich clusters to galaxy groups is
  reproduced only in an `extreme' ejection scheme, where material ejected from
  dark matter haloes is re--incorporated on a timescale comparable to the age
  of the Universe.  Finally, we explore how the metal abundance in the
  intergalactic medium as a function of redshift can constraint how and when
  galaxies ejected their metals.
\end{abstract}


\section[]{Introduction}
\label{sec:intro}

$N$--body simulations have shown that the baryon fraction in a rich
cluster does not change appreciably during its evolution
\citep{white}.  Clusters of galaxies can thus be considered as closed
systems, retaining all information about their past star formation and
metal production histories \citep{renzini}.  This suggests that direct
observations of elemental abundances in the intra--cluster medium
(ICM) can constrain the history of star formation in clusters, the
efficiency with which gas was converted into stars, the relative
importance of different types of supernovae, and the mechanisms
responsible for the ejection and the transport of metals.

The last decade has witnessed the accumulation of a large amount of
data on the chemical composition of the intra--cluster gas
\citep{mushotzky,degrandi,ettori}.  X--ray satellites have provided a
wealth of information about the abundances of many different
elements. These studies have shown that the intra--cluster gas cannot
be entirely of primordial origin -- a significant fraction of this gas
must have been processed in the cluster galaxies and then transported
from the galaxies into the ICM.

The total amount of iron dispersed in the ICM is of the same order of
magnitude as the mass of iron locked in the galaxies
\citep{renzini93}.  Observational data suggest that the mean
metallicity of the ICM is about $0.2$--$0.3$ solar both for nearby
\citep{edge} and for distant clusters \citep{mushotzky2}.

Various physical mechanisms can provide viable explanations for the
transfer of metals from the galaxies into the ICM, for example
ejection of enriched material from mergers of proto--galactic
fragments \citep{gnedin}; tidal/ram pressure stripping
\citep{gunn,fukumoto,mori}; galactic outflows
\citep*{larson,gibson,wiebe}.  The actual contribution from each of
these mechanisms is still a matter of debate.  Observational data
suggest that metals are most effectively transferred into the ICM by
internal mechanisms rather than external forces. \citet{renzini} has
argued that ram pressure cannot play a dominant role, because this
process would operate more efficiently in high velocity dispersion
clusters.  A correlation between the richness of the cluster and its
metal content is not supported by observations.

In recent years, supernova--driven outflows have received increasing
attention as the most plausible explanation for the presence of metals
in the ICM.  It was originally suggested by \citet{larson1} and
\citet{larson} that the fraction of mass and hence of metals driven
from a galaxy increases with decreasing galactic mass, because lower
mass galaxies have shallower potential wells.  This naturally
establishes a metallicity--mass relationship that is in qualitative
agreement with the observations.  It also predicts the chemical
pollution of the ICM as a side--effect of the outflow.

This outflow scenario is not without its problems, however.  It has
been shown in a number of papers \citep*{david,matt,gibson,moretti}
that if a standard IMF and chemical yield is assumed, it is very
difficult to account for the total amount of metals observed in rich
clusters. Some authors have suggested that cluster ellipticals may
form with a non--standard `top--heavy' IMF.  This would alleviate the
metal budget problems and also explain the `tilt' of the fundamental
plane, i.e. the increase in galaxy mass--to--light ratio with
increasing luminosity \citep{zepf,chiosi,pad}.  From a theoretical
point of view, a skewness of the IMF towards more massive stars at
higher redshift might be expected from simple arguments related to the
Jeans scale and to the scale of magnetic support against gravitational
collapse \citep{lars}.  On the other hand, many authors have argued
that there is very little real observational evidence that the IMF
does vary, at least between different regions of our own Galaxy
\citep{hern,massey,kroupa}.  In this analysis, we will simply sweep
the IMF issue under the carpet by treating the chemical yield as one
of the parameters in our model.

Many interesting clues about metal enrichment at high redshift have
been found by studying Lyman break galaxies (LBGs).  Detailed studies
at both optical and infrared wavelengths have shown that at redshifts
$\sim 3$ , the metallicity of LBGs is relatively high (in the range
$0.1$--$0.5$ Z$_{\odot}$) \citep{pettini1,pettini2}.  Studies of the
spectral energy distribution of these objects have shown, somewhat
surprisingly, that $20$ per cent of these galaxies have been forming
stars for more than $1$ Gyr \citep{shapley}.  This pushes the onset of
star formation in these objects to redshifts in excess of $5$.
Although there seems to be a general consensus that star formation
activity (and hence the chemical pollution of the interstellar medium)
must have started at high redshift, the question of \emph{which}
galaxies are responsible for this pollution is still controversial.
Some theoretical studies show that elliptical galaxies must have
played an important role in establishing the observed abundance of the
ICM, but these studies often require, as we noted before, an initial
mass function (IMF) that is skewed towards more massive stars at high
redshift.  Most such modelling has also not been carried out in a
fully cosmological context (see, however, \citet{kauffcharlot} for an
approach closer to that of this paper).  Other studies \citep{garnett}
suggest that dwarf galaxies have been the main contributors to the
chemical pollution of the inter--galactic medium (IGM).

In this paper we use a combination of high--resolution $N$--body
simulations of the formation of clusters in a $\Lambda$CDM Universe
and semi--analytic techniques to follow the enrichment history both of
galaxies and of the ICM.  We test that our model is able to reproduce
observations of the number density, the stellar populations and the
chemical properties of cluster galaxies, as well as the metal content
of the ICM.  We then study which galaxies were primarily responsible
for polluting the ICM and when this occurred.

The paper is structured as follows: in Sec.~\ref{sec:simulations} we
describe the simulations used in this work; in
Sec.~\ref{sec:volkermodel} we summarise the semi--analytic technique
we employ; while in Sec.~\ref{sec:mymodel} we give a detailed
description of the prescriptions adopted to parametrise the physical
processes included in our model.  In Sec.~\ref{sec:results} we
describe how we set the free parameters of our model and we show the
main observational properties that can be fit.
Sec.~\ref{sec:enrichment} and Sec.~\ref{sec:budget} present the main
results of our investigation on the chemical enrichment history of the
ICM and the IGM, and investigate two observational tests that may help
to distinguish between different feedback schemes.  Our conclusions
are presented in Sec.~\ref{sec:conclusions}.


\section[]{$N$--body simulations}
\label{sec:simulations}

In this study we use a collisionless simulation of a cluster of galaxies,
generated using the `zoom' technique \citep*{tormen,katz}.  As a first step, a
suitable target cluster is selected from a previously generated cosmological
simulation.  The particles in the target cluster and its immediate surroundings
are traced back to their Lagrangian region and replaced with a larger number of
lower mass particles.  These particles are then perturbed using the same
fluctuation field as in the parent simulation, but now extended to smaller
scales (reflecting the increase in resolution).  Outside the
\emph{high-resolution} region, particles of variable mass, increasing with
distance, are displaced on a spherical grid whose spacing grows with distance
from the high--resolution region and that extends to the box size of the parent
simulation.  This method allows us to concentrate the computational effort on
the cluster of interest and, at the same time, to maintain a faithful
representation of the large--scale density and velocity of the parent
simulation.  In the following, we will refer to haloes that include
low--resolution particles as `contaminated' haloes.  We will exclude these
haloes in the analysis.

The cluster simulation used in this work was carried out by Barbara Lanzoni as
part of her PhD thesis and is described in \citet{lanzoni} and \citet{me}.  We
also use a re--simulation of a `typical' region of the Universe, carried out by
Felix Stoehr as part of his PhD thesis.  This used the same re--simulation
technique as the cluster simulation.  In both, the fraction of haloes
contaminated by the presence of low--resolution particles is $\sim 3$ per cent,
all near the boundary of the high resolution region.

The parent simulation employed is, in both cases, the Very Large
Simulation (VLS) carried out by the Virgo Consortium
\citep*{jenk,yoshida}.  The simulation was performed using a parallel
P3M code \citep{mf} and followed $512^3$ particles with a particle
mass of $7\times 10^{10}\,h^{-1}\,{\rm M}_{\odot}$ in a comoving box
of size $479\,h^{-1}$Mpc on a side.  The parent cosmological
simulation is characterised by the following parameters:
$\Omega_0=0.3$, $\Omega_{\Lambda}=0.7$, spectral shape $\Gamma=0.21$,
$h=0.7$ (we adopt the convention $H_0=100\,h\,{\rm km}\,{\rm
s}^{-1}\,{\rm Mpc}^{-1}$) and spectral normalisation $\sigma_8=0.9$.

The numerical parameters of the simulations used in this work are
summarised in Table~\ref{tab:nums}.

\begin{table*}
\caption{Numerical parameters for the simulations used. Both the
  simulations were carried out assuming a $\Lambda$CDM cosmology with
  cosmological parameters $\Omega_0=0.3$, $\Omega_{\Lambda}=0.7$,
  $\Gamma=0.21$, $\sigma_8=0.9$, and $h=0.7$. In the table, we give the
  particle mass $m_{\rm p}$ in the high resolution region, the
  starting redshift $z_{\rm start}$ of the simulation, and the 
  gravitational softening $\epsilon$ in the high--resolution region.}

\begin{tabular}{lllll}
\hline
Name & Description & $m_{\rm p}$ [$h^{-1}$M$_{\odot}$] & $z_{\rm start}$ &
$\epsilon$ [$h^{-1}$kpc]  \\
\hline

g$1$ & $10^{15}\,h^{-1}{\rm M}_{\odot}$ cluster & $2.0\times 10^9$ & 60 & 5.0\\

M$2$ & field simulation & $9.5 \times 10^8$ & 70 & 3.0\\
\hline
\end{tabular}
\label{tab:nums}
\end{table*}


\section[]{Tracking Galaxies in $N$--body Simulations}
\label{sec:volkermodel}

The prescriptions adopted for the different physical processes
included in our model are described in more detail in the next
section. In this section we summarise how the semi--analytic model is
grafted onto the high resolution $N$--body simulation.  The techniques
we employ in this work are similar to those used by \citet{volker}.

In standard semi--analytic models, all galaxies are located within dark matter
haloes.  Haloes are usually identified in a simulation using a standard
friends--of--friends (FOF) algorithm with a linking length of $0.2$ in units of
the mean particle separation.  The novelty of the analysis technique developed
by Springel et al., is that substructure is also tracked within each halo.
This means that the dark matter halo within which a galaxy forms, is still
followed even after it is accreted by a larger object.  The algorithm used to
identify subhaloes ({\small SUBFIND}) is described in detail by \citet{volker}.
The algorithm decomposes a given halo into a set of disjoint and self--bound
subhaloes, identified as locally overdense regions in the density field of the
background halo.  \citet{me} have presented an extensive analysis of the
properties of the subhalo population present in a large sample of haloes with a
range of different masses.  As in \citet{me}, we consider all substructures
detected by the {\small SUBFIND} algorithm with at least $10$ self-bound
particles, to be genuine subhaloes.

An important change due to the inclusion of subhaloes, is a new
nomenclature for the different kinds of galaxies present in the
simulation.  The FOF group hosts the `central galaxy'; this galaxy is
located at the position of the most bound particle in the halo.  This
galaxy is fed by gas cooling from the surrounding hot halo medium.
All other galaxies attached to subhaloes are called `halo galaxies'.
These galaxies were previously central galaxies of another halo, which
then merged to form the larger object.  Because the core of the parent
halo is still intact, the positions and velocities of these halo
galaxies can be accurately determined.  Note that gas is no longer
able to cool onto halo galaxies.

Dark matter subhaloes lose mass and are eventually destroyed as a
result of tidal stripping effects.  A galaxy that is no longer
identified with a subhalo is called a satellite.  The position of the
satellite is tracked using the position of the most bound particle of
the subhalo before it was disrupted.  Note that if two or more
subhaloes merge, the halo galaxy of the smaller subhalo will become a
`satellite' of the remnant subhalo.

\citet{volker} show that the inclusion of subhaloes results in a
significant improvement in the cluster luminosity function over
previous semi--analytic schemes, and in a morphology--radius
relationship that is in remarkably good agreement with the
observational data.  This improvement is mainly attributed to a more
realistic estimate of the merger rate: in the standard scheme too many
bright galaxies merge with the central galaxy on short time--scales.
This produces first--ranked galaxies that are too bright when compared
with observational data.  It also depletes the cluster luminosity
function around the `knee' at the characteristic luminosity.


\section[]{The physical processes governing galaxy evolution}
\label{sec:mymodel}

Our treatment of the physical processes driving galaxy evolution is
similar to the one adopted in \citet{k99} and \citet{volker}.  Many
prescriptions have been modified in order to properly take into
account the exchange of metals between the different phases.  We also
include metallicity--dependent cooling rates and luminosities.  The
details of our implementation are described below.  The reader is
referred to previous papers for more general information on
semi--analytic techniques
\citep*{whitefrenk,k93,baugh,k99,somerville,cole}.

\subsection[]{Gas cooling}
\label{sec:cooling}

Gas cooling is treated as in \citet{k99} and \citet{volker}.  It is
assumed that the hot gas within dark matter haloes initially follows
the dark matter distribution. The cooling radius is defined as the
radius for which the local cooling time is equal to age of the
Universe at that epoch.
  
At early times and for low--mass haloes, the cooling radius can be
larger than the virial radius. It is then assumed that the hot gas
condenses out on a halo dynamical time.  If the cooling radius lies
within the virial radius, the gas is assumed to cool quasi--statically
and the cooling rate is modelled by a simple inflow equation.

Note that the cooling rates are strongly dependent on the temperature
of the gas and on its metallicity.  We model these dependences using
the collisional ionisation cooling curves of \citet{sutherland}.  At
high ($\ge 10^8$ K) temperatures, the cooling is dominated by the
bremsstrahlung continuum.  At lower temperatures line cooling from
heavy elements dominates (mainly iron in the $10^6$--$10^7$ K regime,
with oxygen significant at lower temperatures).  The net effect of
using metallicity--dependent cooling rates is an overall increase of
the brightness of galaxies, because cooling is more efficient.  This
effect is strongest in low mass haloes.

As noted since the work of \citet{whitefrenk}, these prescriptions
produce central cluster galaxies that are too massive and too luminous
to be consistent with observations.  This is a manifestation of the
`cooling flow' problem, the fact that the central gas in clusters does
not appear to be cooling despite the short estimated cooling time.  As
in previous models \citep{k99,volker} we fix this problem \emph{ad
hoc} by assuming that the gas does not cool in haloes with $V_{\rm
vir} > V_{\rm cut}$.  In our model $V_{\rm cut}=350\,{\rm km}\,{\rm
s}^{-1}$.

Note that following \citet{volker}, we define the virial radius $R_{\rm vir}$
of a FOF-halo as the radius of the sphere centred on its most-bound particle
which has an overdensity $200$ with respect to the critical density.  We take
the enclosed mass $M_{\rm vir} = 100 H^2 R^3_{\rm vir}/G$ as the virial mass,
and we define the virial velocity as $V^2_{\rm vir}= G M_{\rm vir}/R_{\rm
  vir}$.  The mass of a subhalo, on the other hand, is defined in terms of the
total number of particles it contains.  The virial velocity of a subhalo is
fixed at the velocity that it had just before infall.

Although our cooling model is extremely simplified, it has been shown
that it produces results that are in good agreement with more detailed
$N$--body $+$ hydrodynamical simulations that adopt the same physics
\citep{naoki,helly}.

\subsection[]{Star formation}
\label{sec:starformation}

The star formation `recipes' that are implemented in semi--analytic
models are always subject to considerable uncertainty.  In \citet{k99}
and \citet{volker} it is assumed that star formation occurs with a
rate given by:
\begin{equation}
\label{eq:sfr}
\psi = \alpha M_{\rm cold}/t_{\rm dyn}
\end{equation}
where $M_{\rm cold}$ and $t_{\rm dyn} = R_{\rm vir}/10V_{\rm vir}$ are
the cold gas mass and the dynamical time of the galaxy respectively,
and $\alpha$ represents the efficiency of the conversion of gas into
stars.

Previous semi--analytic models have assumed that the efficiency
parameter $\alpha$ is a constant, independent of galaxy mass and
redshift.  There are observational indications, however, that low mass
galaxies convert gas into stars less efficiently than high mass
galaxies \citep{k02}.  An effect in this direction is also expected in
detailed models of the effects of supernovae feedback on the
interstellar medium \citep{mck,efsta}.  More importantly, perhaps, the
above prescription leads to gas fractions that are essentially
independent of the mass of the galaxy.  In practice, we know that gas
fractions increase from $\sim 0.1$ for luminous spirals like our own
Milky Way, to more than $0.8$ for low--mass irregular galaxies
\citep{boissier}.

In this work, we assume that $\alpha$ depends on the circular velocity
of the parent galaxy as follows:
\begin{displaymath}
\alpha = \alpha_0 \cdot \left(\frac{V_{\rm vir}}{220\,{\rm km}\,{\rm 
    s}^{-1}}\right)^n 
\end{displaymath}
and we treat $\alpha_0$ and $n$ as free parameters.  Note also that
$R_{\rm vir}$ decreases with redshift for a galaxy halo with fixed
circular velocity.  This means that a galaxy of circular velocity
$V_c$ will be smaller at higher redshifts and as a result, the star
formation efficiency will be higher.

\subsection[]{Feedback}
\label{sec:feedback}

Previous work \citep{kauffcharlot,somerville,cole} has shown that
feedback processes are required to fit the faint end slope of the
luminosity function and to fit the observed slope of the
colour--magnitude relation of elliptical galaxies.  The theoretical
and observational understanding of how the feedback process operates
is far from complete.

In many models it has been assumed that the feedback energy released
in the star formation process is able to reheat some of the cold gas.
The amount of reheated mass is computed using energy conservation
arguments and is given by:
\begin{equation}
\label{eq:feed}
\Delta M_{\rm reheated} = \frac{4}{3} \epsilon \frac{\eta_{\rm SN} E_{\rm
    SN}}{V^{2}_{\rm vir}} \Delta M_{\rm star}
\end{equation}
where $\eta_{\rm SN}$ is the number of supernovae expected per solar
mass of stars formed ($6.3\times 10^{-3} {\rm M}^{-1}_{\odot}$
assuming a universal \citet{salpeter} IMF) and $E_{\rm SN}$ is the
energy released by each supernova ($\simeq 10^{51}$ erg).  The
dimensionless parameter $\epsilon$ quantifies the efficiency of the
process and is treated as a free parameter.  Note that changing the
IMF would also change the amount of energy available for reheating the
gas.

One major uncertainty is whether the reheated gas leaves the halo.
This will depend on a number of factors, including the velocity to
which the gas is accelerated, the amount of intervening gas, the
fraction of energy lost by radiative processes, and the depth of the
potential well of the halo.

On the observational side, evidence in support of the existence of
outflows from galaxies has grown rapidly in the last years
\citep{heckman,marlowe,martin}.  In many cases, the observed gas
velocities exceed the escape velocity of the parent galaxies; this
material will then escape from the galaxies and will be injected into
the intergalactic medium (IGM).  Observations of galactic--scale
outflows of gas in active star--forming galaxies
\citep*{lehne,dahle,heckman00,tim} suggest that outflows of multiphase
material are ubiquitous in galaxies in which the global
star--formation rate per unit area exceeds roughly $10^{-1}{\rm
M}_{\odot}\,{\rm yr}^{-1}\,{\rm kpc}^{-2}$.  Different methods to
estimate the outflow rate suggest that it is comparable to the star
formation rate.  The estimated outflow speeds vary in the range
$400$--$800\,{\rm km}\,{\rm s}^{-1}$ and are independent of the galaxy
circular velocity.  These observational results indicate that the
outflows preferentially occur in smaller galaxies. This provides a
natural explanation for the observed relation between galaxy
luminosity and metallicity.

An accurate implementation of the feedback process is beyond the
capabilities of present numerical codes.  As a result, published
simulation results offer little indication of appropriate recipes for
treating galactic winds.  In this paper we experiment with three
different simplified prescriptions for feedback and study whether they
lead to different observational signatures:
\begin{itemize}
\item in the {\em retention} model, we use the prescriptions adopted by
  \citet{kauffcharlot} and assume that the reheated material, computed
  according to Eq.~\ref{eq:feed}, is shock heated to the virial
  temperature of the dark halo and is put directly in the hot phase,
  where it is then once more available for cooling.
\item In the {\em ejection} model, we assume that the material
  reheated by supernovae explosions in central galaxies always leaves
  the halo, but can be later re--incorporated. The time--scale to
  re--incorporate the gas is related to the dynamical time--scale of
  the halo by the following equation:
  \begin{equation}
    \label{eq:back}
    \Delta M_{\rm back} = \gamma \cdot M_{\rm ejected} \cdot 
    \frac{ V_{\rm vir}}{ R_{\rm vir}} \cdot \Delta t 
  \end{equation}
  where $\Delta M_{\rm back}$ is the amount of gas that is
  re--incorporated in the time--interval $\Delta t$; $M_{\rm ejected}$
  is the amount of material in the ejected component; $R_{\rm vir}$
  and $V_{\rm vir}$ are the virial radius and the virial velocity of
  the halo at the time the re--incorporation occurs; $\gamma$ is a
  free parameter that controls how rapidly the ejected material is
  re--incorporated. Note that the material ejected in this way is not
  available for cooling until it is re--incorporated in the hot
  component.
  
  For all the other galaxies (halo and satellite galaxies), we assume
  that the material reheated to the virial temperature of the subhalo
  is then kinematically stripped and added to the hot component of the
  main halo.
\item In the {\em wind} model, we adopt prescriptions that are
  motivated by the observational results.  We assume that only central
  galaxies residing in haloes with a virial velocity less that $V_{\rm
  crit}$ can eject outside the halo. The outflow rates from these
  galaxies are assumed to be proportional to their star formation
  rates, namely:
  \begin{equation}
    \label{eq:eqwind}
    \dot{M}_{\rm w} = c \cdot \psi
  \end{equation}
  Observational studies give values for $c$ in the range $1$--$5$
  \citep{martin2} while values in the range $100$--$300$ ${\rm
  km}\,{\rm s}^{-1}$ are reasonable for $V_{\rm crit}$ \citep{tim}.
  We treat $V_{\rm crit}$ and $c$ as free parameters. We also assume
  that the ejected material is re--incorporated as in the ejection
  scheme.  If the conditions for an outflow are not satisfied, the
  reheated material (computed according to Eq.~\ref{eq:feed}) is
  treated in the same way as in the retention model.  The wind model
  is thus intermediate between the retention and the ejection schemes.
  Satellite galaxies are treated as in the ejection scheme.
\end{itemize}

\subsection[]{Galaxy mergers}
\label{sec:mergers}

In hierarchical models of galaxy formation, galaxies and their
associated dark matter haloes form through merging and accretion.
          
In our high resolution simulations, mergers between subhaloes are
followed explicitly.  Once a galaxy is stripped of its dark halo,
merging timescales are estimated using a simple dynamical friction
formula: \citep{bt}
\begin{displaymath}
T_{\rm friction} = \frac{1}{2} \frac{f(\epsilon)}{C} \frac{V_{\rm vir} R_{\rm
    vir}^2}{GM_{\rm sat} {\rm ln}\Lambda} 
\end{displaymath}

\citet*{navarro} show that this analytic estimate is a good fit to the results
of numerical simulations.  The formula applies to satellites of mass $M_{\rm
  sat}$ orbiting at a radius $R_{\rm vir}$ in a halo of virial velocity $V_{\rm
  vir}$.  $f(\epsilon)$ expresses the dependence of the decay on the
eccentricity of the orbit and is well approximated by $f(\epsilon) \sim
\epsilon^{0.78}$ \citep{lc}; $C$ is a constant $\sim 0.43$ and ${\rm
  ln}\Lambda$ is the Coulomb logarithm.  We adopt the average value
$<f(\epsilon)> \sim 0.5$, computed by \citet{bepi}, and approximate the Coulomb
logarithm with ${\rm ln}\Lambda = (1+M_{\rm vir}/M_{\rm sat})$.  For the
satellite galaxy mass we use the value of $M_{\rm vir}$ corresponding at the
last time the galaxy was a central galaxy (either of a halo or of a subhalo).

When a small satellite merges with the central galaxy, its stellar mass and
cold mass are simply transferred to the central galaxy and the photometric
properties are updated accordingly.  In particular we transfer the stellar mass
of the merged galaxy to the bulge of the central galaxy and update the
photometric properties of this galaxy.  If the mass ratio between the stellar
component of the merging galaxies is larger than $0.3$, we assume that the
merger completely destroys the disk of the central galaxy producing a
spheroidal component.  In addition we assume that the merger consumes all the
gas left in the two merging galaxies in a single burst. The stars formed in
this burst are also added to the bulge.  Note that since the galaxy is fed by a
cooling flow, it can grow a new disk later on.

\subsection[]{Spectro--photometric evolution}
\label{sec:lum}

The photometric properties of our model galaxies are calculated using
the models of \citet{bc}, which include the effect of metallicity on
the predicted luminosities and colours of a galaxy.  The stellar
population synthesis models are used to generate look--up tables of
the luminosity of a single burst of fixed mass, as a function of the
age of the stellar population and as a function of its metallicity.
When updating the photometric properties of our model galaxies, we
interpolate between these tables using a linear interpolation in $t$
and ${\rm log}\,Z$.  It is assumed that stars form with the same
metallicity as the cold gas.  We have adopted a \citet{salpeter} IMF
with upper and lower mass cut--offs of $100$ and $0.1\,{\rm
M}_{\odot}$.

\citet*{charlot} have demonstrated that for a given IMF and star
formation history, the broad--band colours produced by different
stellar population codes differ by only a few tenths of a magnitude.
The most important sources of uncertainty in our model predictions are
thus the IMF (which affects the luminosity quite strongly) and the
associated yield (which influences the colours).

\subsection[]{Dust extinction}
\label{sec:dust}

Attenuation of starlight by dust affects the colours of galaxies.  The
properties of dust are dependent on a number of factors such as the star
formation rate, that regulates the rate of creation, heating and destruction of
dust grains and the distribution of dust and metals within gas clouds.  For a
single galaxy, all these factors can be taken into account, and it is then
possible to model the effect of dust on the galaxy's spectrum.  However, the
level of detail that is required goes far beyond the capabilities of our
present code.  We therefore adopt a dust model that is based on the macroscopic
properties of galaxies, i.e. luminosity and inclination.

This model has been used in previous work \citep{k99,somerville} and is based
on the observational results by \citet{wang}, who studied the correlation
between the face--on optical depth of dust in galactic discs and the total
luminosity of the galaxy. \citet{wang} find that this can be expressed as:
\begin{displaymath}
\tau_{\rm B} = \tau_{{\rm B},*}  \left(\frac{L_{\rm B}}{L_{{\rm B},
    *}}\right)^{\beta} 
\end{displaymath} 
\noindent
where $L_{\rm B}$ is the intrinsic (unextincted) blue luminosity and $L_{{\rm
    B},*}$ is the fiducial observed blue luminosity of a Schechter $L_*$ galaxy
$(M_*({\rm B}) = -19.6 + 5 {\rm log} h)$. \citet{wang} find that a relation
with $\tau_{{\rm B},*} = 0.8 \pm 0.3$ and $\beta \sim 0.5$ fit their data very
well.

We use $\tau_{{\rm B},*} = 0.8$, $\beta = 0.5$ and relate the B--band optical
depth to the other bands using the extinction curve of \citet*{cardelli}. We
also assign a random inclination to each galaxy and apply the dust correction
only to its disc component (i.e. we assume that the bulge is not affected by
dust) using a `slab' geometry:
\begin{displaymath}
A_{\lambda} = -2.5 {\rm log} 
\frac{1-e^{-\tau_{\lambda}sec\theta}}{\tau_{\lambda}sec\theta}
\end{displaymath}
\noindent
where $\theta$ is the angle of inclination to the line of sight.  All the
magnitudes and the colours plotted in this paper include the effects of dust
extinction, unless stated otherwise.

\subsection[]{Metal routes}
\label{sec:routes}

We assume that a yield Y of heavy elements is produced per solar mass
of gas converted into stars.  All the metals are instantaneously
returned to the cold phase (note that this means that we are assuming
a mixing efficiency of $100$ per cent).  We also assume that a
fraction (R) of the mass in stars is returned to the cold gas.

Metals are then exchanged between the different gas phases depending
on the feedback model (see Sec.~\ref{sec:feedback}).  In the retention
model, the metals contained in the reheated gas are put in the hot
component and can subsequently cool to further enrich the cold phase.
In the ejection and wind schemes, the metals can be ejected outside
the halo. The ejected metals are re--incorporated into the hot halo
gas on a halo dynamical time (see Eq.~\ref{eq:back}).  In the wind
scheme, metals are only ejected for haloes with $V_{\rm vir} < V_{\rm
crit}$.  In more massive systems, the metals contained in the reheated
gas are mixed with the hot component of the main halo.  Some
simulations \citep{ferrara} suggest that low--mass galaxies may lose
essentially all their metals, but it is difficult for these galaxies
to `blow--away' their interstellar medium.  For simplicity, we do not
consider schemes in which metals are selectively ejected from
galaxies.

When a satellite galaxy merges, its stars, cold gas and metals are
simply added to those of the central galaxy.  If a \emph{major} merger
occurs, all the gas is consumed in a starburst and the metals are
ejected into the hot phase in the retention scheme and into the
ejected phase in the other two schemes (in the wind scheme ejection
occurs only in galaxies that reside in haloes with $V_{\rm vir} >
V_{\rm crit}$).

\begin{figure}
\bc
\resizebox{8cm}{!}{\includegraphics{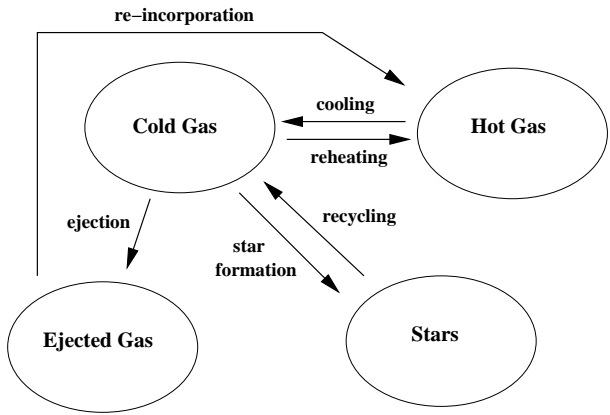}}\\%
\caption{A schematic representation of how mass is exchanged between the
  different phases considered in our models in the absence of accretion from
  outside.  Each arrow is accompanied by a name indicating the physical process
  driving the mass exchange. Metals follow the same routes as the mass.}
\label{fig:scheme}
\ec
\end{figure}

In Fig.~\ref{fig:scheme} we sketch the routes whereby mass and metals
are exchanged in the model.  We now write down the equations that
describe the evolution of the mass of the four reservoirs shown in
Fig.~\ref{fig:scheme}.  All central galaxies have $4$ different
components: stars, cold gas, hot gas and an `ejected' component.  The
equations describing these $4$ components (in the absence of accretion
of external matter) are:\\ ~\\
\noindent
$\dot{M}_{\rm stars} = (1 - {\rm R}) \cdot \psi$  \\
~\\
$\dot{M}_{\rm hot} = -\dot{M}_{\rm cool} + \dot{M}_{\rm back} +
\sum_{{\rm sat}} \dot{M}_{\rm reheated}$\\
~\\
$\dot{M}_{\rm cold} = + \dot{M}_{\rm cool} - (1 - {\rm R}) \cdot \psi - 
\dot{M}_{\rm out}$
\\ 
~\\
$\dot{M}_{\rm ejected} = + \dot{M}_{\rm out} - \dot{M}_{\rm back}$\\ 
~\\
\noindent
where $\dot{M}_{\rm out}$ is the rate at which the cold gas is ejected
outside the halo and is given by Eq.~\ref{eq:feed} for the ejection
scheme and Eq.~\ref{eq:eqwind} for the wind scheme; $\psi$ is the star
formation rate given by Eq.~\ref{eq:sfr}; $\dot{M}_{\rm cool}$ is the
cooling rate; $\dot{M}_{\rm back}$ is the rate at which gas is
reincorporated from the ejected material (Eq.~\ref{eq:back});
$\sum_{\rm sat} \dot{M}_{\rm reheated}$ is the sum of all material
that is reheated by the satellite galaxies in the halo.

Satellite galaxies have no hot gas or ejected components and
$\dot{M}_{\rm cool} =0$.  The equations for the satellite galaxies
are:\\ ~\\
\noindent
$\dot{M}_{\rm stars} = (1 - {\rm R}) \cdot \psi$  \\
~\\
$\dot{M}_{\rm cold} = + \dot{M}_{\rm cool} - (1 - {\rm R}) \cdot \psi - 
\dot{M}_{\rm reheated}$\\
~\\
\noindent
where $\dot{M}_{\rm reheated}$ is computed using Eq.~\ref{eq:feed}.

From the above equations, one can easily obtain the corresponding
equations for the evolution of the metal content of each reservoir of
material.  For central galaxies, the equations are:\\ ~\\
\noindent
$\dot{M}^Z_{\rm stars} = + (1 - {\rm R}) \cdot \psi \cdot Z_{\rm cold}$\\
~\\
$\dot{M}^Z_{\rm hot} =  - \dot{M}_{\rm cool} \cdot Z_{\rm hot} +
\dot{M}_{\rm back} \cdot Z_{\rm ejected} + \sum_{{\rm sat}} [\dot{M}_{\rm
  reheated} \cdot Z_{\rm cold}]$\\ 
~\\
$\dot{M}^Z_{\rm cold} = + \dot{M}_{\rm cool} \cdot Z_{\rm hot} - (1 - {\rm R}) 
\cdot \psi \cdot Z_{\rm cold} + {\rm Y} \cdot \psi - \dot{M}_{\rm out} 
\cdot Z_{\rm cold}$\\   
~\\
$\dot{M}^Z_{\rm ejected} = +  \dot{M}_{\rm out} \cdot Z_{\rm cold} - 
\dot{M}_{\rm back} \cdot Z_{\rm ejected}$\\ 
~\\
\noindent 
where $Z_{\rm cold} = M^Z_{\rm cold}/M_{\rm cold}$, $Z_{\rm ejected} =
M^Z_{\rm ejected}/M_{\rm ejected}$ and $Z_{\rm hot} = M^Z_{\rm
hot}/M_{\rm hot}$ represent the metallicities in the cold gas phase,
in the ejected component and in the hot gas phase respectively.

Note that the above equations assume that material is being ejected
outside haloes.  In our `retention' scheme, no material leaves the
halo and the applicable equations are obtained by neglecting the
equation for $\dot{M}_{\rm ejected}$ and substituting $\dot{M}_{\rm
back}$ and $\dot{M}_{\rm out}$ with $\dot{M}_{\rm reheated}$ (given by
Eq.~\ref{eq:feed}).
 
The practical implementation of the prescriptions is described in
detail in \citet{volker}.  We have stored $100$ outputs of the
simulation spaced in equal logarithmic intervals in redshifts from
$z=20$ to $z=0$.  For each new snapshot, we estimate the merger
timescales of satellite galaxies that have entered a given halo.  The
`merger clock' for the other satellites is updated and a galaxy merges
with the central object when this time has elapsed.  The merger
clock is reset if the halo containing the satellites merges with a
larger system.

The total amount of hot gas available for cooling in each halo is
given by:
\begin{equation}
\label{eq:hotmass}
  M_{\rm hot} = f_{\rm b}M_{\rm vir}-\sum_i [M_{\rm star}^{(i)}+
  M_{\rm cold}^{(i)}+M_{\rm ejected}^{(i)}]
\end{equation}
where the sum extends over all the galaxies in the halo and $f_{\rm
b}$ is the baryon fraction of the Universe.  In this work we use the
value $f_{\rm b}=0.15$ as suggested by the CMB experiment WMAP
\citep{spergel}.  When going from one snapshot to the next, three
things may happen:
\begin{itemize}
\item a new halo may form.  In this case its hot gas mass is initialised to the
  value $f_{\rm b}M_{\rm vir}$;
\item the virial mass of the halo may increase because of accretion of
  `diffuse' material.  In this case the accreted baryons are effectively added
  to the hot component using Eq.~\ref{eq:hotmass};
\item two haloes can merge.  In this case both the ejected component and the
  hot component of the lower mass halo are added to the hot component of the
  remnant halo.  The ejected component of the remnant halo remains outside the
  halo but is reduced by a factor given by Eq.~\ref{eq:back}.
\end{itemize}

The cooling rate is assumed to be constant between two successive
simulation outputs.  The differential equations given above are solved
using smaller time--steps ($50$ between each pair of simulation
snapshots).
 
\subsection[]{Model normalisation and influence of the parameters}
\label{sec:params}

As in previous work \citep{kauffcharlot,somerville}, the free parameters of the
model are tuned in order to reproduce the observed properties of our Galaxy
(see Table~\ref{tab:MW}).  We also check that we get the right normalisation
and slope for the Tully--Fisher relation.

Note that the main parameters controlling the physical evolution of our model
galaxies are:
\begin{itemize}
\item the parameters determining the star formation efficiency: $\alpha_0$ and
  $n$;
\item the parameters determining the feedback efficiency: $\epsilon$ in the
  ejection model, and $c$ and $V_{\rm crit}$ in the wind model;
\item the amount of metals produced per solar mass of cold gas converted into
  stars: Y;
\item the gas fraction returned by evolved stars: R.
\item the parameter $\gamma$ that determines how long it takes for the ejected
  gas to be re--incorporated.
\end{itemize}

The value of the parameter R can be directly estimated from stellar evolution
theory for a given choice of IMF.  Population synthesis models show that this
recycled fraction is roughly independent of metallicity and lies in the range
$0.2$--$0.45$ \citep{cole}.  Note that this value can be larger for top--heavy
IMFs.

Note that there are only a few free parameters in our model.  The influence of
the different parameters on the observed properties of galaxies can be
summarised as follows:
\begin{itemize}
\item the yield Y controls the total amount of metals in the stars and gas.
  Note that the cooling rates in lower mass haloes are strongly dependent on
  the metallicity.  Increasing the yield results in an overall increase of the
  luminosities of our galaxies and a tilt in the Tully--Fisher relation towards
  a shallower slope.
\item The star formation efficiency $\alpha_0$ has only a weak influence on the
  zero--point of the Tully--Fisher relation, but it has an important influence
  on the gas fraction of galaxies.
\item The parameter $n$, that parametrises the dependence of star formation
  efficiency on the circular velocity of the galaxy, has a strong effect on the
  dependence of the gas fraction on the mass or circular velocity of the
  galaxy.  It has negligible effect on all the other observational properties.
\item The feedback efficiency has a strong influence both on the zero--point
  and the slope of the Tully--Fisher relation.  An increase in feedback
  efficiency results in a decrease in the luminosities of galaxies and a tilt
  in the Tully--Fisher towards a steeper slope.
\item The gas fraction returned by evolved stars $R$ has only a marginal
  influence on the gas metallicity and on the luminosities of the galaxies in
  our model.
\item The parameter $\gamma$ controls how long it takes for the ejected gas to
  be re--incorporated back into a dark matter halo. We have experimented with
  $\gamma$ values in the range $0.1$--$1$. If $\gamma=0.1$, the
  re--incorporation time is of order the Hubble time. If $\gamma$ is large, the
  ejection model simply reverts back to the retention scheme. Note that if we
  decrease $\gamma$, gas and metals remain outside dark matter haloes longer
  and cooling rates are reduced.  As a result, the feedback efficiency in our
  preferred model is smaller.  In the next section, we will show results for
  $\gamma=0.1$, which should be considered an `extreme' ejection scheme, where
  gas and metals remain outside the halo for a time comparable to the age of
  the Universe.  In Sec.~\ref{sec:budget}, we will explore in more detail what
  happens if the re--incorporation timescale is reduced.
\end{itemize} 

\begin{table}
\bc
 \caption{Free parameters values adopted for the three models.}
 \begin{tabular}{@{}lrrrrrrr@{}}
           & $\alpha_0$ & $n$ & $\epsilon$& Y   & R&  $c$   & $V_{\rm crit}$\\
 retention & $0.09$     & $2.2$ &  $0.45$  & $0.045$& $0.35$ & --    & --\\
 ejection  & $0.08$     & $2.2$ &  $0.15$  & $0.040$& $0.35$ & --    & --\\
 wind      & $0.10$     & $2.5$ &  $0.35$  & $0.040$& $0.35$ & $5$   & $150$
 \end{tabular}
\label{tab:parameters}
\ec
\end{table}

A through exploration of parameter space results in the parameters listed in
Table \ref{tab:parameters}.  The requirement that our models agree with a wide
range of observational data (see Sec.~\ref{sec:results}) allows only slight
changes around the values listed in Table \ref{tab:parameters}.

Note that in all three models, we have to assume a value for the yield that is
larger than the conventional value.  This is consistent with other analyses,
which have shown that the observed metallicity of the ICM cannot be explained
using a standard IMF with a standard value of the yield \citep*{gibson,gib}. A
recent review of the problem can be found in \citet{moretti}.  In this work, we
will leave aside this problem.  Note also that we will not attempt to
distinguish between the heavy elements produced by SNII and SNIa.  It is known
that the latter are the most important contributors for Fe, while the former
mainly contribute $\alpha$ elements.  SNII and SNIa events are characterised by
different time--scales, so that there is a lag between the ejection of these
elements into the interstellar medium.  This will be studied in more detail in
a future paper (Cora et al., in preparation).  We will also not attempt to
explore what would happen if the various free parameters listed in
Table~\ref{tab:parameters} were to depend on redshift.  A redshift dependence
of the star formation efficiency, for example, may be required to explain the
evolution in the number density of luminous quasars \citep{kauffh} and to
reproduce the observed properties of Lyman break galaxies at redshift $\sim 3$
\citep*{somm}.

Note that the only parameter that changes significantly between the three
different models is the feedback efficiency and that for the retention and wind
schemes, we are obliged to adopt an uncomfortably high value.  The value
adopted for the parameter $c$ also lies on the upper end of the allowed range.
Efficient feedback is required in order to counteract the high cooling rates
and prevent overly luminous galaxies from forming at the present day.  If the
ejected material is kept outside the haloes for substantial periods, then the
need for high feedback efficiencies is not as great.  Note that a high value of
the feedback efficiency is also required in the wind model.  This is because
most of the galaxies in the simulation box rapidly fall below the conditions
required for a `wind', given the relatively low values adopted for $V_{\rm
  crit}$.  Efficient feedback is then again required in order to avoid the
formation of overly luminous galaxies.

\section[]{Comparison with local observations}
\label{sec:results}

In this section, we present model results for some of the basic observed
properties of galaxies, both in and out of clusters.  As explained in
Sec.~\ref{sec:params}, the free parameters of our model are mainly tuned in
order to reproduce the observed properties of the Milky--Way and the correct
normalisation of the Tully--Fisher relation .  An extensive exploration of
parameter space seems to indicate that the range of acceptable parameters for
each model is small.

We select as `Milky--Way type' galaxies all the objects in the simulation with
circular velocities in the range $200$--$240\,{\rm km}\,{\rm s}^{-1}$ and with
bulge--to--disk ratios consistent with Sb/Sc type galaxies \citep{simien}.
More specifically we select all galaxies with $1.5 < \Delta M < 2.6$ ($\Delta
M= M_{\rm bulge} - M_{\rm total}$).  Note that in our model we assume that the
circular velocity of a galaxy is $\sim 25$ per cent larger than the circular
velocity of its halo. This is motivated by detailed models \citep*{mo} for the
structure of disk galaxies embedded in cold dark matter haloes with the
universal NFW profile \citep*{nfw}.  These models show that the rotation
velocity measured at twice the scale length of the disk is $20$--$30$ per cent
larger than the virial velocity of the halo.

In our three models we find $11$, $13$ and $11$ Milky--Way type galaxies (for
the retention, ejection and wind model respectively) that reside in
uncontaminated haloes.  For these galaxies we obtain the stellar masses, gas
masses, star formation rates and metallicities listed in Table~\ref{tab:MW}.
These values are very close to what is observed for our own Galaxy, which has a
stellar mass $\sim 10^{11}\,{\rm M}_{\odot}$ and a total mass of cold gas in
the range $6.5\cdot10^9$--$5.1\cdot10^{10}\,{\rm M}_{\odot}$
\citep{k99,somerville}.  \citet{rochapinto1,rochapinto2} give an estimate for
the mean SFR in the Milky Way's disc over the last few Gyr of the order of
$1$--$3\,{\rm M}_{\odot}\,{\rm yr}^{-1}$.  This is somewhat lower than the
values we find for our model galaxies.  The Galaxy has a B--band and I--band
absolute magnitude of $\sim -20.5$ and $\sim -22.1$ respectively, and a
V--light weighted mean metallicity of $\sim 0.7$ solar
\citep{kauffcharlot,somerville}.  All these observed values are in reasonably
good agreement with the values listed in Table~\ref{tab:MW}.

\begin{table*}
 \centering
 \caption{Properties of \emph{Milky--Way} type galaxy in our simulation. The 
   masses are in units of $h^{-1}\,{\rm M}_{\odot}$ and the SFR is in units of
   ${\rm M}_{\odot}\,yr^{-1}$.}
 \begin{tabular}{@{}lccccccc@{}}
           & $M_{\rm gas}$ & $M_{\rm star}$ & SFR & $Z_{\rm stars}/Z_{\odot}$ &
             $M_B - 5 log(h)$     & $M_V - 5 log(h)$   & $M_I - 5 log(h)$\\
  \hline

  retention& $8.03\cdot10^{9}$    & $4.67\cdot10^{10}$  & $4.57$& $0.80$ &
             $-20.49$             & $-21.07$           & $-22.08$\\

  ejection & $7.53\cdot10^{9}$   & $4.94\cdot10^{10}$  & $3.76$& $0.82$ &
             $-20.42$             & $-21.03$           & $-22.07$\\


  wind     & $8.83\cdot10^{9}$   & $6.00\cdot10^{10}$  & $5.25$& $0.87$ &
             $-20.70$             & $-21.27$           & $-22.30$\\

\hline
 \end{tabular}\\
\begin{flushleft}
\end{flushleft}
\label{tab:MW}
\end{table*}

\begin{figure}
\bc
\resizebox{8cm}{!}{\includegraphics{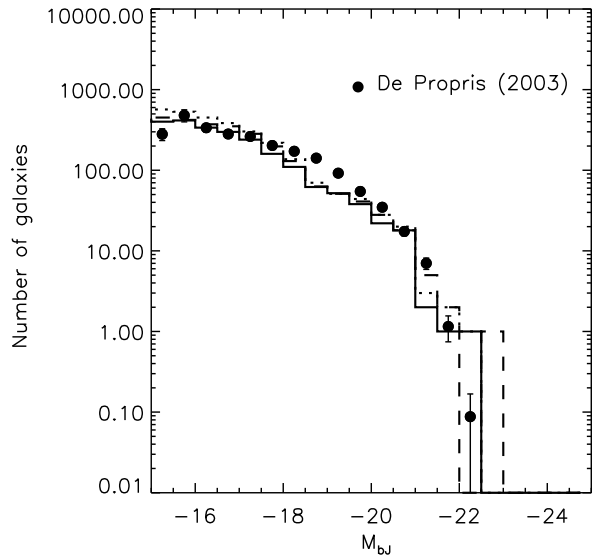}}\\%
\caption{Luminosity function in the b$_{\rm J}$--band for model galaxies in 
  the retention, ejection and wind scheme.  The points represent the composite
  luminosity function by \citet{depropris}.  The solid line is for the
  retention model, the dashed line for the ejection model and the dotted line
  for the wind model.}
\label{fig:LF}
\ec
\end{figure}

In Fig.~\ref{fig:LF} we show cluster luminosity functions in the b$_{\rm
  J}$--band for each of our three schemes.  We find cluster luminosity
functions which are in a reasonably good agreement with the composite
luminosity function for cluster galaxies obtained by \citet{depropris}. A fit
to a \citet{schec} function gives a characteristic magnitude of $M_{b_{\rm
    J}}^*=-21.47$ and a faint--end power law slope of $\alpha=-1.22$ for the
retention model. Note that resolution effects artificially flatten the
luminosity function at magnitudes $\gtrsim -15$ (the fit to a Schechter
function is performed on the magnitude interval from $-16$ to $-22$).  For the
ejection and wind schemes the results are similar.  The cluster galaxies are
slightly more luminous and the faint--end slope is somewhat steeper.

\begin{figure}
\bc
\resizebox{8.5cm}{!}{\includegraphics{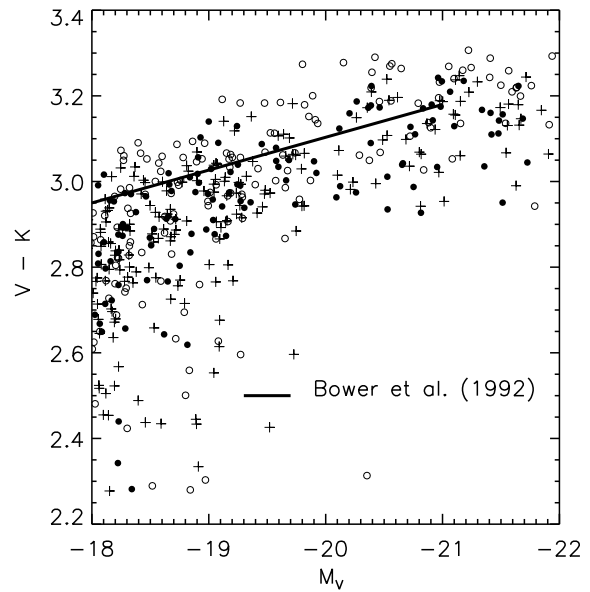}}\\%
\caption{Colour--magnitude relation for early type galaxies in the simulations
  compared with the observational relation obtained for Coma cluster
  ellipticals by \citet{bower}. Filled circles are for the retention model,
  empty circles for the ejection model and crosses for the wind model.}
\label{fig:CM}
\ec
\end{figure}

Fig.~\ref{fig:CM} shows that our model galaxies correctly reproduce the slope
of the observed colour--magnitude relation for cluster ellipticals.  The solid
line is the fit to data for Coma cluster from \citet*{bower}.  The points are
model galaxies with an early type morphology ($\Delta M < 0.2$) for our three
different schemes.  Our results confirm the conclusion that the
colour--magnitude is mainly driven by metallicity effects
\citep{kodama,kauffcharlot}.  Note that the scatter of the model galaxies is
larger than the observational value: for all three model we find a scatter of
$\sim 0.07$, while the value measured by \citet{bower} is $0.048$.

\begin{figure}
\bc
\resizebox{8.5cm}{!}{\includegraphics{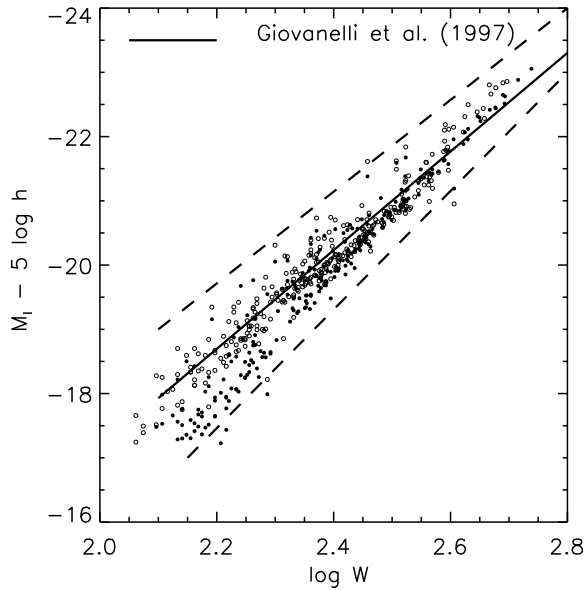}}\\%
\caption{Tully--Fisher relation for our model galaxies compared with
  the observational result by \citet{giovanelli}.  The scatter in the
  observational sample is shown with dashed lines while the thick
  solid line shows the `best fit' relation given by Giovanelli et
  al. The filled circles represent model galaxies in the ejection
  model while the empty circles are used the model galaxies from the
  wind model.}
\label{fig:TF}
\ec
\end{figure}

In Fig.~\ref{fig:TF} we show the Tully--Fisher relation obtained for our model
galaxies.  We have plotted central galaxies that are in haloes outside the main
cluster and that are not contaminated by low resolution particles.  We apply a
morphological cut that selects Sb/Sc--type galaxies ($1.5 < \Delta M < 2.6$)
and we select all galaxies brighter than $-18$ in the I--band.  These
morphological and magnitude cuts correspond approximately to the ones defining
the sample used by \citet{giovanelli}. The mean observational relation is shown
by a solid line in the figure, and the scatter about the relation is indicated
by the dashed lines.  The relation of Giovanelli et al. is already corrected
for internal extinction.  We therefore do not correct our I--band magnitudes
for dust extinction in this plot.

Note that the Tully--Fisher relations in our simulations have extremely small
scatter.  There are probably many sources of additional scatter that we do not
account for -- for example photometric errors and scatter in the relation
between halo circular velocity and observed line--widths.  The slope of our
predicted TF relation is also somewhat steeper than the observations for the
retention and ejection models (for clarity only model galaxies from the
ejection scheme are plotted as filled circles in the figure; galaxies in the
retention scheme exhibit a relation that is very similar).  We obtain the best
fit to the observed Tully--Fisher relation for our wind model (shown as empty
circles).  This is because in this scheme the ejected mass is systematically
smaller than in the ejection scheme for low mass galaxies.  The slope of the
Tully--Fisher relation is strongly dependent on the adopted feedback
prescriptions and could in principle be used to test different feedback models
if we had better control of the systematic effects in converting from
theoretical to observed quantities.


\begin{figure}
\bc
\resizebox{8.5cm}{!}{\includegraphics{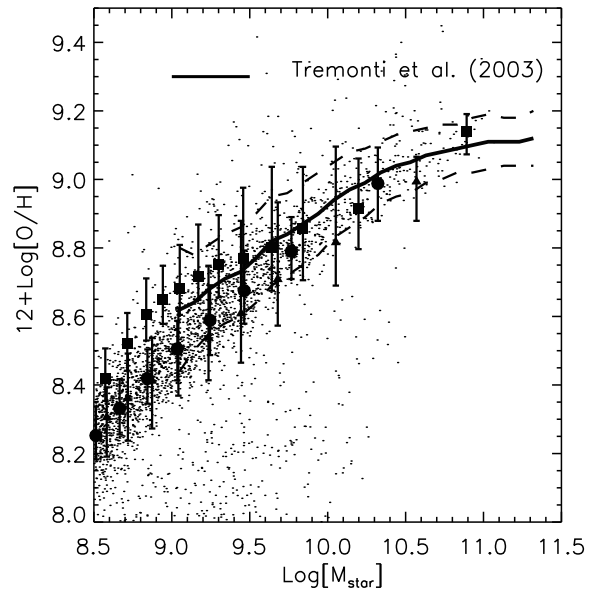}}\\
\caption{Metallicity--mass relation for our model galaxies (shown as
  points).  The thick solid line represents the median relation
  between stellar mass and metallicity obtained for a sample of
  $50,000$ galaxies in the SDSS (Tremonti et al., in preparation). The
  dashed lines indicate the scatter in the observed relation.  The
  solid symbols represent the median obtained from our models in bins
  containing $\sim 400$ galaxies each. Filled circles are for the
  retention model, filled triangles are for the ejection model and
  filled squares are for the wind model. The error bars mark the
  $20{\rm th}$ and $80{\rm th}$ percentiles of the distribution.}
\label{fig:ZvsMass}
\ec
\end{figure}

In Fig.~\ref{fig:ZvsMass} we compare the relation between metallicity and
stellar mass for our simulated galaxies with the stellar mass--metallicity
relation derived from new data from the Sloan Digital Sky Survey (Tremonti et
al., in preparation).  Measuring the metallicities of stars in a galaxy is
difficult, because most stellar absorption lines in the spectrum of a galaxy
are sensitive to both the age of the stellar population and its metallicity.
The metallicity of the gas in a galaxy can be measured using strong emission
lines such as [OII], [OIII], [NII], [SII] and H$\beta$, but up to now, the
available samples have been small.  Tremonti et al. have measured gas--phase
metallicities for $\sim 50,000$ emission--line galaxies in the SDSS and the
median relation from their analysis is plotted as a solid line in
Fig.~\ref{fig:ZvsMass}.  Note that the stellar masses for the SDSS galaxies are
measured by Tremonti et al.  using the same method as in \citet{k02}.  The
dashed lines represent the scatter in the observed relation.  The dots in the
figure represent simulated galaxies in the retention model. We have only
selected galaxies that reside in uncontaminated haloes outside the main cluster
and that have gas fractions of at least $10$ per cent.  All of these galaxies
are star--forming and should thus be representative of an emission
line--selected sample.
We obtain very similar relations for our two other schemes.  This can
be seen from the solid symbols in the plot, which indicate the median
in bins that each contain $\sim 400$ model galaxies.  Note that
for the retention and the ejection scheme, the model metallicities are
sistematically lower than the observed values, although within
$1\sigma$ from the median observed relation.  However recent work
suggests that strong line methods (as used in Tremonti et al.) may
systematically overestimate oxygen abundances by as much as
$0.2$--$0.5$ dex \citep*{ken}.  Given the uncertainties in both the
stellar mass and the metallicity measurements, the agreement between
our models and the observational results is remarkably good.
 
\begin{figure}
\bc
\resizebox{8.5cm}{!}{\includegraphics{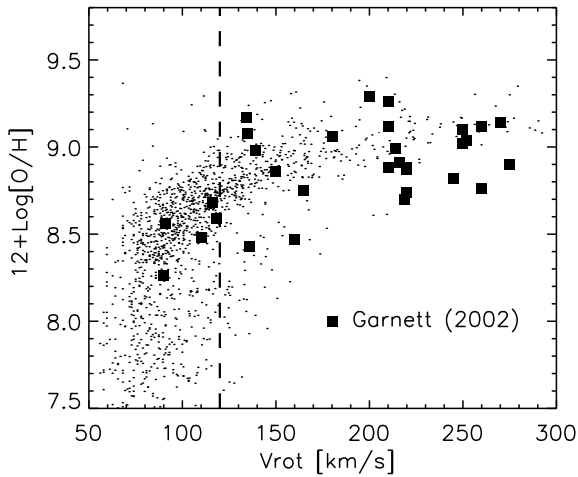}}\\%
\caption{Relation between metallicity and rotational velocity for our
  model galaxies (shown as points). The filled squares represent the
  observational data from \citet{garnett}. The vertical dashed line in
  the panel corresponds to a velocity $120\,{\rm km}\,{\rm s}^{-1}$
  (the turn--over velocity claimed by Garnett).}
\label{fig:Garnett}
\ec
\end{figure}

In Fig.~\ref{fig:Garnett} we compare the metallicity versus rotational velocity
measurements published by \citet{garnett} to the results obtained for our
models. For simplicity, we only show results from the retention model. The
other two schemes give very similar answers.  Again we have selected only
galaxies that reside in uncontaminated haloes outside the main cluster and that
have gas fractions of at least $10$ per cent.  The vertical dashed line in the
figure shows a velocity of $120\,{\rm km}\,{\rm s}^{-1}$. According to the
analysis of Garnett, this marks a threshold below which there is a stronger
dependence of metallicity on the potential well depth of the galaxy.  Note,
however, that the plot shown by Garnett is in linear units in $V_{\rm rot}$ and
that the `turnover' in the mass--metallicity relation is much more convincing
in the data of Tremonti et al.  As can be seen, our models fit the
metallicity--rotational velocity data as well.  This is not surprising, given
that we obtain a reasonably good fit to the observed Tully--Fisher relation.

\begin{figure}
\bc
\resizebox{8.5cm}{!}{\includegraphics{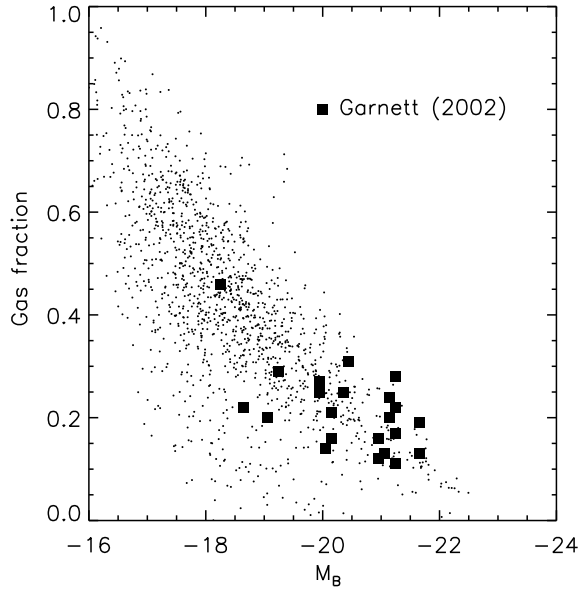}}\\%
\caption{Gas fraction as function of the B--band luminosity for our
  model galaxies (shown as points). Filled squares are the gas
  fractions computed by \citet{garnett} for the same sample of data
  shown in Fig.~\ref{fig:Garnett}.  Results are shown for the
  retention model, but are very similar for the other two schemes.}
\label{fig:gasfrac}
\ec
\end{figure}

In Fig.~\ref{fig:gasfrac} we show a comparison between the gas
fraction of galaxies in our models and the gas fractions computed by
\citet{garnett}. The same sample of objects as in
Fig.~\ref{fig:Garnett} is plotted, both for the models and for the
observations.  Again our model agrees well with the observational
data.

\section[]{Chemical enrichment of the ICM}
\label{sec:enrichment}

In our simulated cluster, the metallicity in the hot gas component is $\simeq
0.26$--$0.30~Z_{\odot}$ which is in good agreement with X--ray measurements
\citep{renzini93}.

As pointed out by Renzini et al., simple metal abundances in the ICM depend not
only on the total amount of metals produced in stars, but also on how much
dilution there has been from pristine gas.  A quantity that is not dependent on
this effect or on the total mass in dark matter in the cluster is the
so--called iron mass--to--light ratio (IMLR), defined as the ratio between the
mass of iron in the ICM and the total B-band luminosity of cluster galaxies.

Our simulated cluster has a total mass of $1.14\cdot10^{15}\,h^{-1}\,{\rm
  M}_{\odot}$ and a total luminosity in the B band of $\sim
10^{13}\,L_{\odot}$.  Assuming a solar iron mass fraction from
\citet*{grevesse}, we find IMLR $=0.015-0.020\,{\rm M}_{\odot}\,L^{-1}_{\odot}$
for our three models, in agreement with the range given by \citet{renzini93}.

Table \ref{tab:ML} lists the total mass--to--light ratios of our
cluster in the V and B--bands. The results are given for the three
different feedback schemes.  Note that the observational determination
of a cluster mass--to--light ratio is not an easy task.  Both the
mass and the luminosity estimates are affected by uncertainties.
Estimates based on the virial mass estimator give $\Upsilon_{\rm V}
\sim 175$--$252$ \citep{carlberg,girardi} while B--band
mass--to--light ratios are in the range
$200$--$400\,h^{-1}\,\Upsilon_{\odot}$ \citep{kent,girardi2}. It has
been argued that the virial mass estimator can give spurious results
if substructure is present in the cluster or if the volume sampled
does not extend out to the virial radius.  In general, estimates based
on masses derived from X--ray data tend to give lower values.

\begin{table}
 \centering
 \caption{Mass--to--light ratios in the V and in the B--band for our
   three models. The units are in $h\,\Upsilon_{\odot}$.}
 \begin{tabular}{@{}lrr@{}}
           & $\Upsilon_{\rm B}$ & $\Upsilon_{\rm V}$ \\
 retention & $290$            & $230$            \\
 ejection  & $250$            & $200$            \\
 wind      & $240$            & $190$  
 \end{tabular}
\label{tab:ML}
\end{table}

The good agreement between the model predictions and the observational results
indicates that our simulation may provide a reasonable description of the
circulation of metals between the different baryonic components in the
Universe.  We now use our models to generate a set of {\em predictions} for
when the metals in the ICM were ejected and for which galaxies were primarily
responsible for the chemical pollution.

Recall that we assume that metals are recycled instantaneously and that the
chemical pollution of the ICM happens through two routes:
\begin{itemize}
\item in the retention scheme the reheated mass (along with its metals) is
  ejected directly into the hot component.  In this model, the enrichment of
  the ICM occurs at the same time as the star formation.
\item In the ejection scheme the reheated material (along with its metals)
  stays for some time outside the halo and is later re--incorporated into the
  hot component.  This means that there is a delay in the enrichment of the ICM
  in this scheme.
\item The wind scheme sits somewhere in between the retention and ejection
  models.  Metals are ejected out of the halo by galaxies that satisfy the
  outflow conditions. Otherwise, the metals are ejected directly into the hot
  gas.
\end{itemize}

\begin{figure}
\bc
\resizebox{8.7cm}{!}{\includegraphics{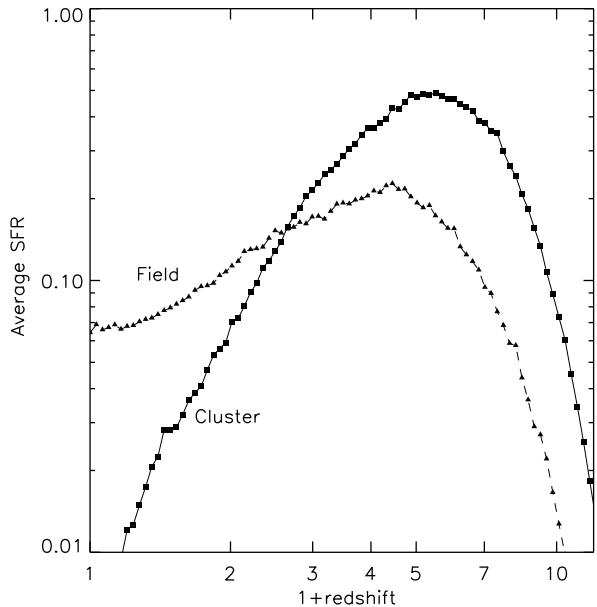}}\\%
\caption{Average star formation rate as function of redshift in the cluster 
  (solid line with filled squares) and in the field (dashed line with filled
  triangles).}
\label{fig:sfr}
\ec
\end{figure}

The instantaneous recycling assumption means that the epoch of the production
of the bulk of metals in the ICM coincides with the epoch of the production of
the bulk of stars.  In Fig.~\ref{fig:sfr} we show the \emph{average} star
formation rate (SFR) of galaxies that end--up in the cluster region and the
average star formation rate for galaxies that end--up in the field (the field
is defined as consisting of all haloes outside the main cluster that are not
contaminated by low resolution particles).

The SFRs in the two regions are normalised to the total amount of stars formed.
The figure shows that star formation in the cluster is peaked at high redshift
($\sim 5$) and drops rapidly after redshift $3$.  The peak of the SFR in the
field occurs at lower redshift ($\sim 3$). The decline in star formation from
the peak to the present day is also very much shallower in the field.

Integrating the SFR history of the cluster, we find that $\sim 85$ per cent of
the stars in the cluster formed at redshift larger than $2$ and $\sim 70$ per
cent at redshift larger than $3$.  The corresponding values in the field are
$60$ per cent and $40$ per cent.  The results are similar in all the three
models.

As explained in Sec.~\ref{sec:intro}, there is no general consensus on
which galaxies were responsible for enriching the ICM.  Several
theoretical studies have proposed that mass loss from elliptical
galaxies is an effective mechanism for explaining the observed amount
of metals in the ICM.  Some of these studies require an IMF that is
skewed towards more massive stars at high redshift
\citep{vettolani,gibson,moretti}.  \citet{kauffcharlot} used
semi--analytic models to model the enrichment of the ICM and found
that a significant fraction of metals come from galaxies with circular
velocities less than $125\,{\rm km}\,{\rm s}^{-1}$. There have also
been a number of more recent attempts to address this question in
using hydrodynamic simulations \citep{aguirre,sh}.

In a recent observational study, \citet{garnett} has analysed the dependence of
the so--called `effective yield' on circular velocity for a sample of irregular
and spiral galaxies (the data are compared to our model galaxies in
Fig.~\ref{fig:Garnett}), concluding, as \citet{larson1} had earlier, that
galaxies with $V_{\rm rot} \le 100-150\,{\rm km}\,{\rm s}^{-1}$ lose a large
fraction of their SN ejecta, while galaxies above this limit retain most of
their metals. As we have noted previously, this observational sample is limited
both in number and in dynamic range.

\begin{figure}
\bc
\resizebox{8.7cm}{!}{\includegraphics{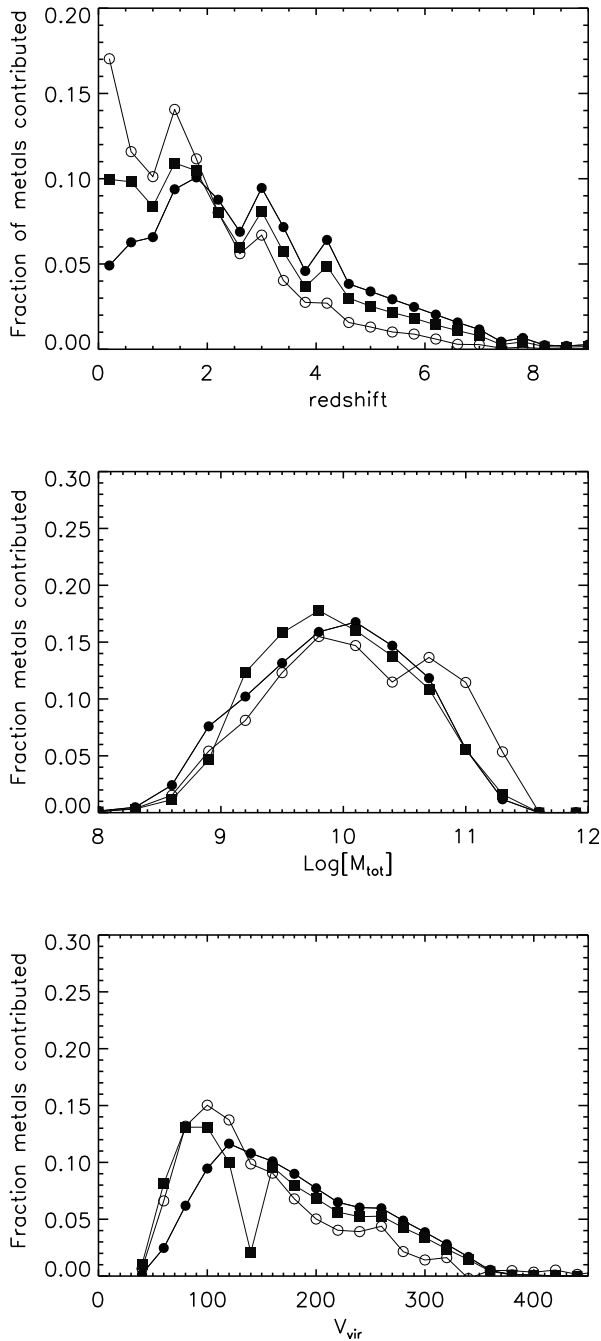}}\\%
\caption{Fraction of metals, today present in the ICM, as a function of the
  redshift they were first incorporated into the intra--cluster gas (upper
  panel), as a function of the total mass of the ejecting galaxy (middle panel)
  and as a function of the virial velocity of the parent substructure (lower
  panel).  Filled circles are used for retention model, empty circles are used
  for ejection model and filled squares are used for wind model.}
\label{fig:ICM}
\ec
\end{figure}

Our model predictions are given in Fig.~\ref{fig:ICM}.  The top panel shows the
fraction of metals in the cluster at present day as a function of redshift they
were first incorporated into the ICM.  The other two panels show the fraction
of metals as a function of the baryonic mass (middle) and the circular velocity
(bottom) of the ejecting galaxies.

The expected delay in the chemical enrichment of the ICM is evident if one
compares the ejection and wind models with the retention model in the top panel
of the figure.  Note that a similar delay is also seen in the metallicity of
the hot component although it is very small because the reduction in total
accreted ICM almost balances that in accreted metals. Over the redshift range
$0$--$2$ the difference between the metallicity of the hot component in the
retention and ejection schemes is only $1.5$ per cent solar.  As expected, the
behaviour of the wind model is intermediate between the ejection and retention
schemes.  We find that $60$--$80$ per cent of the metals are incorporated into
the ICM at redshifts larger than $1$, $35$--$60$ per cent at redshifts larger
than $2$ and $20$--$45$ per cent at redshifts larger than $3$ (we obtain lower
values for the ejection scheme, higher values for the retention scheme and
intermediate values for the wind model).

The middle panel shows that $43$--$52$ per cent of the metals are ejected by
galaxies with total baryonic mass less than $1\times10^{10}\,h^{-1}{\rm
  M}_{\odot}$.  The distribution of the masses of the ejecting galaxies is
approximatively independent of the feedback scheme.  Although low mass galaxies
dominate the luminosity function in terms of number, they do not dominate the
contribution in mass.  Approximately half of the contribution to the chemical
pollution of the ICM is from galaxies with baryonic masses larger than
$1\times10^{10}\,h^{-1}{\rm M}_{\odot}$.

In our model, the star formation and feedback processes depend on the virial
velocity of the parent substructure. In the bottom panel, we show the metal
fraction as a function of the virial velocity of the ejecting galaxies.  We
find that $80$--$88$ per cent of the metals were ejected by galaxies with
virial velocities less than $250\,{\rm km}\,{\rm s}^{-1}$.  The `dip' visible
for the wind model in this panel corresponds to the sharp value of $V_{\rm
  crit}$ we adopt for this feedback scheme.  Note that the results we find from
this analysis are very similar to the results found by \citet{kauffcharlot},
even though the modelling details are different.

\section[]{Observational tests for different feedback schemes}
\label{sec:budget}

Our analysis has shown that properties of galaxies at low redshifts are rather
insensitive to the adopted feedback scheme after our free parameters have been
adjusted to obtain a suitable overall normalisation.
         
The observational tests that would clearly distinguish between the different
models are those that are sensitive to the amount of gas or metals that have
been ejected outside dark matter haloes.

X--ray observations directly constrain the amount of hot gas in massive
virialized haloes. The gas fraction tends to decrease as the X--ray temperature
of the system goes down. \citet*{djf} estimated that the gas--to--total mass
fraction decreases by a factor $\sim 2$--$3$ from rich clusters to groups.  In
elliptical galaxies, the hot gas fraction is ten times lower than in rich
clusters.  These results were confirmed by \citet{sand} in recent study of $66$
clusters and groups with X--ray data.  One caveat is that the hot gas in groups
is detected to a much smaller fraction of the virial radius than in rich
clusters, so it is not clear whether current estimates accurately reflect
global gas fractions \citep{low}.  Another argument for why the gas fraction
must decrease in galaxy groups comes from constraints from the soft X--ray
background.  If galaxy groups had the same gas fractions as clusters, the
observed X--ray background at $0.25$ keV would exceed the observed upper limit
by an order of magnitude \citep*{pen,wu}.

\begin{figure*}
\bc
\resizebox{19cm}{!}{\includegraphics{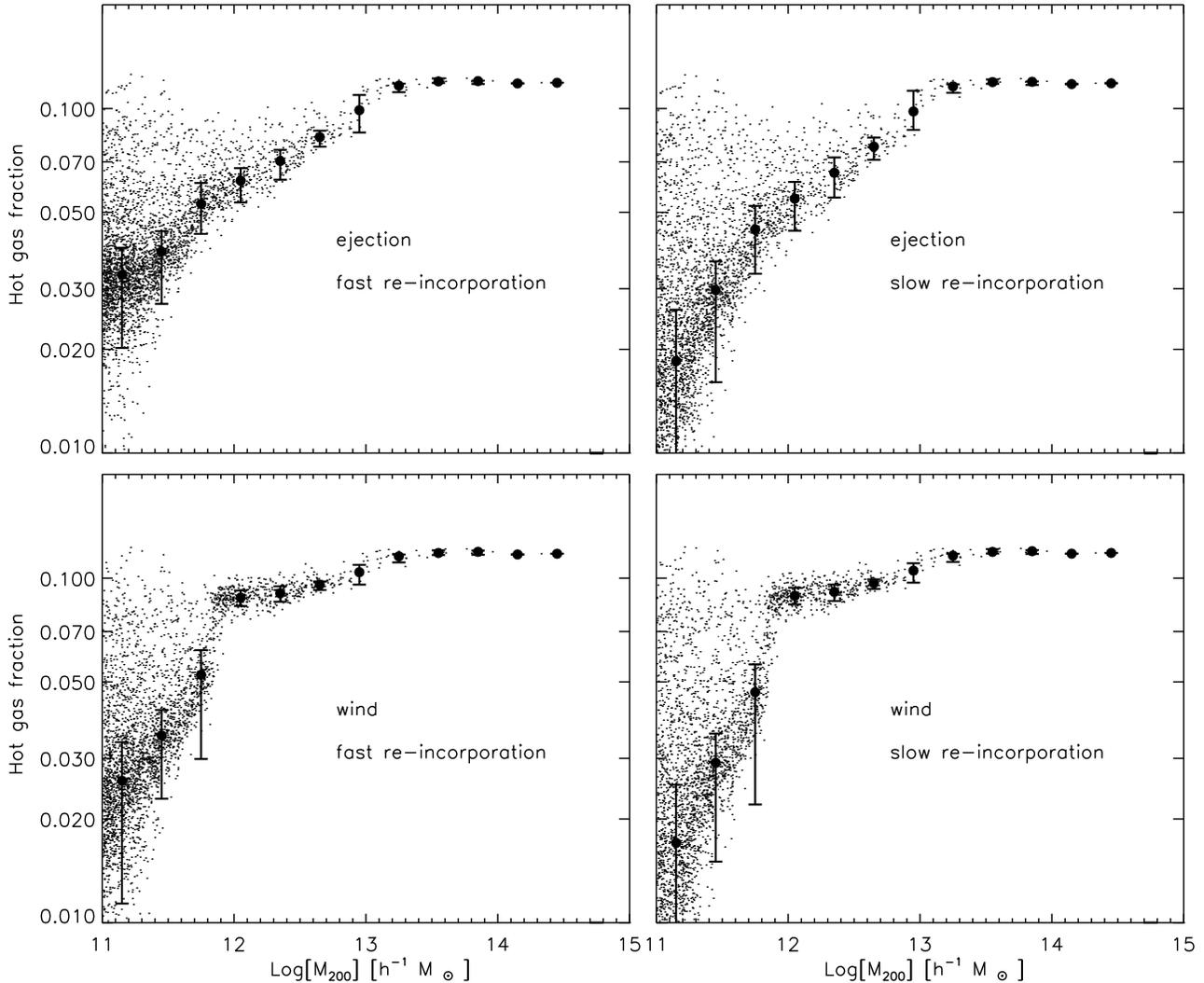}}\\%
\caption{Fraction of hot gas as a function of $M_{200}$.  Solid
  circles represent the median and the error bars mark the $20{\rm
  th}$ and $80{\rm th}$ percentiles of the distributions.}
\label{fig:baryonfrac}
\ec
\end{figure*}

In Fig.~\ref{fig:baryonfrac} we plot the gas fraction, i.e.  $M_{\rm
  hot}/(M_{\rm hot}+M_{\rm vir})$, as a function of virial mass for all the
haloes in our simulation. The mass of the hot gas is computed using
Eq.~\ref{eq:hotmass}. The results are shown for the case of a \emph{short}
re--incorporation timescale ($\gamma = 0.3$) in the left column and for a
\emph{long} re--incorporation timescale ($\gamma = 0.1$) in the right column.
In this figure we plot results for the M2 simulation (see
Table~\ref{tab:nums}). Note that when we change $\gamma$, we also have to
adjust the feedback efficiency $\epsilon$ so that the model has the same
overall normalisation in terms of the total mass in stars formed (for $\gamma =
0.3$, we assume $\epsilon = 0.2$).  We find that the gas fraction remains
approximately constant for haloes with masses comparable to those of clusters,
but drops sharply below masses of $10^{13} M_{\odot}$. As expected, the drop is
more pronounced for the model with a long re-incorporation timescale.

The trends for the wind scheme are similar to those for the ejection scheme,
but as can be seen, the `break' towards lower gas fractions occurs at lower
halo masses, because no material is ejected for haloes with circular velocities
larger than $V_{\rm crit}$. In the ejection scheme, material is ejected for
haloes up to a circular velocity of $350\,{\rm km}\,{\rm s}^{-1}$.  Above this
value, cooling flows shut down and no stars form in the central galaxies.
Galaxies continue, however, to fall into the cluster.  When a satellite galaxy
is accreted, we have assumed that its ejected component is re--incorporated
into the hot ICM.  It is this infall of satellites that causes the gas
fractions to saturate in all schemes at halo masses larger than a few times
$10^{13} M_{\odot}$.

The ejection model with a long re--incorporation timescale is perhaps closest
to satisfying current observational constraints. It reproduces the factor
$2$--$3$ drop in gas fraction between rich clusters and groups. By the time one
reaches haloes of $10^{12} M_{\odot}$, the gas fractions have decreased to a
few percent.  This may help explain why there has so far been a failure to
detect any diffuse X-ray emission from haloes around late-type spiral galaxies
\citep{benson,kuntz}.  Note that in the retention model, we have assumed that
the material reheated by supernovae explosions never leaves the halo.  This
translates into a hot gas fraction that is almost constant as a function of
$M_{200}$.  This model is therefore not consistent with the observed decrease
in baryon fraction from rich clusters to galaxy groups.

Another possible way of constraining our different feedback schemes is to study
what fraction of the metals reside in the diffuse intergalactic medium, well
away from galaxies and their associated haloes.  We now study how metals are
partitioned among stars, cold gas, hot halo gas and the `ejected component' and
we show how this evolves as a function of redshift.

We plot the evolution of the metal content in the different components for our
three different feedback schemes. For the ejection scheme we show results for
two different re-incorporation timescales.  The metal mass in each phase is
normalised to the total mass of metals in the simulation at redshift zero.
This is simply the yield Y multiplied by the total mass in stars formed by
$z=0$.  Note that in the retention scheme, all the reheated gas is put into the
halo and there is no ejected component.

\begin{figure}
\bc
\resizebox{8.5cm}{!}{\includegraphics{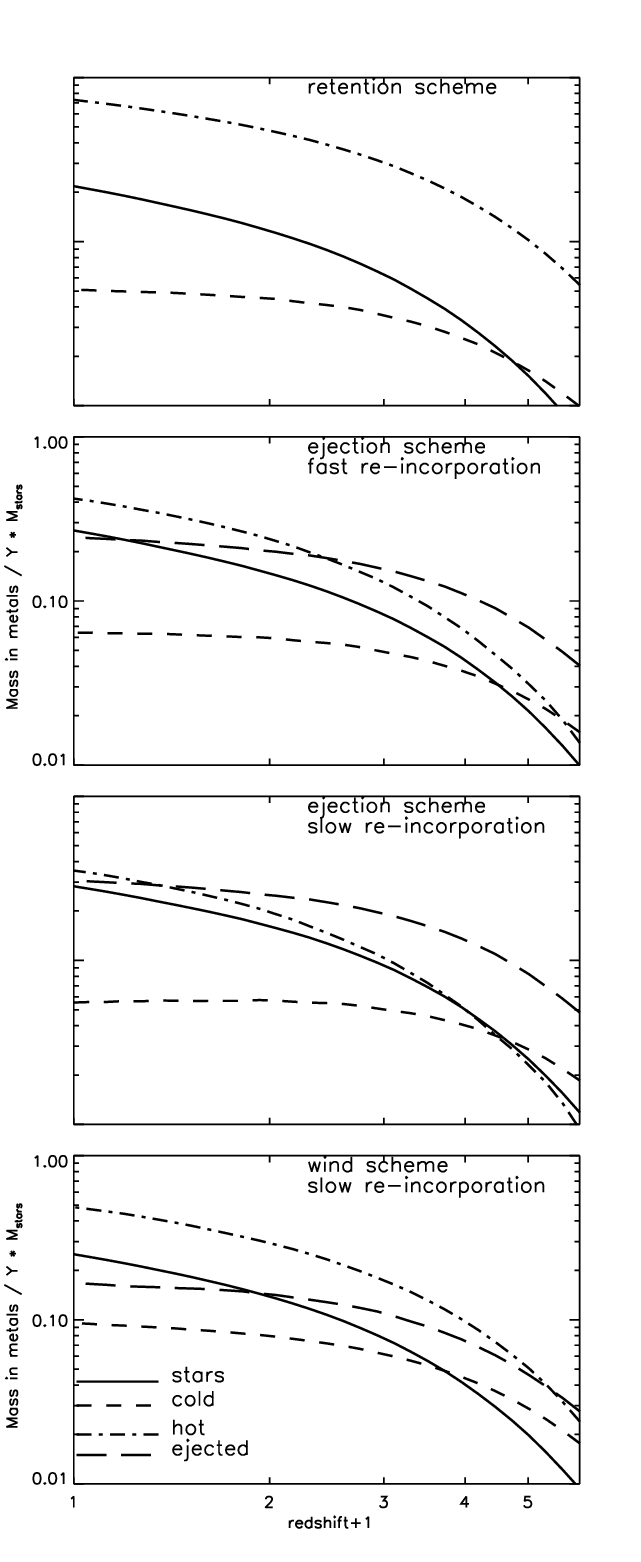}}\\%
\caption{Evolution of the metal content of different phases for the three 
  different models used in this paper and for a typical `field' region.  In
  each panel the solid line represents the evolution of the metal content in
  the stars, the dashed line the cold gas, the dashed--dotted line the hot gas
  and the long--dashed line the ejected component (not present in the retention
  model).  The metal content in each phase is normalised to the total mass of
  metals produced from all the galaxies considered.}
\label{fig:budget}
\ec
\end{figure}

\begin{figure}
\bc
\resizebox{8.5cm}{!}{\includegraphics{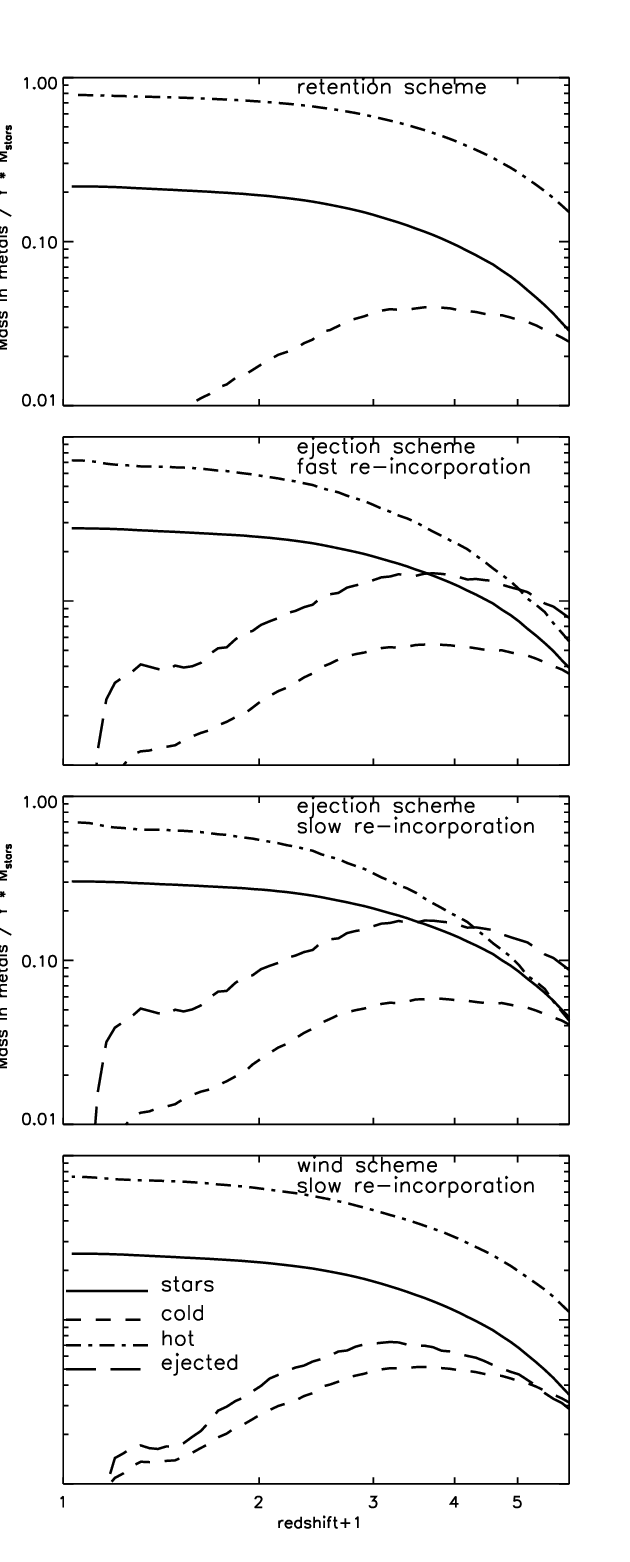}}\\%
\caption{Same as in Fig.~\ref{fig:budget} but for the galaxies within
  the virial radius of our cluster simulation.  Different lines have
  the same meaning as in Fig.~\ref{fig:budget}.}
\label{fig:budgetcl}
\ec
\end{figure}

Fig.~\ref{fig:budget} shows the evolution of the metallicity for an average
`field' region of the Universe (our simulation M2).  In contrast,
Fig.~\ref{fig:budgetcl} shows results for our cluster.  In this plot we only
consider the metals ejected from galaxies that reside within the virial radius
of the cluster at the present day.  We also normalise to the total mass of
metals inside this radius, rather than the total mass of metals in the whole
box. Because very little material is ejected from massive clusters in any of
our feedback schemes, the ejected component always falls to zero by $z=0$ in
Fig.~\ref{fig:budgetcl}.

Let us first consider the field simulation (Fig.~\ref{fig:budget}).  In the
retention scheme, more than $70$ per cent of the metals are contained in the
hot gas.  About $20$--$30$ per cent of the metals are locked into stars and
around $10$ per cent of the metals are in cold gas in galaxies.  In the
ejection scheme, a large fraction of the metals reside outside dark matter
haloes.  This is particularly true for the `slow' re-incorporation scheme,
where the amounts of metals in stars, in hot gas and in the ejected component
are almost equal at $z=0$. For the `fast' re-incorporation model, the metals in
the hot gas still dominate the budget at low redshifts.

In the wind scheme, the amount of metals outside virialized haloes is
considerably lower. There are a factor $2$ more metals in the hot gas than
there are in stars at $z=0$.

At higher redshifts, the relative fraction of metals contained in cold galactic
gas and in the ejected component increases. This is because galaxies are less
massive and reside in dark matter haloes with lower circular velocities.  As a
result, they have higher gas fractions (see Sec.~\ref{sec:starformation}) and
are able to eject metals more efficiently.  In the ejection scheme, the
material that is reheated by supernovae is always ejected from the halo,
irrespective of the circular velocity of the system. In this scheme, metals in
the ejected component (i.e. in diffuse intergalactic medium) dominate at
redshifts greater than 2 in the case of fast re-incorporation and at all
redshifts in the case of slow re-incorporation.  In the wind scheme, material
is only ejected if the circular velocity of the halo lies below some critical
threshold.  In this scheme, there are always more metals associated with dark
matter haloes than there are in the ejected component.

Turning now to the cluster (Fig.~\ref{fig:budgetcl}), we see that the ejected
component is negligible in all three schemes.  The amount of metals outside
dark matter haloes never exceeds the amount of metals locked up in stars up to
redshift $\sim 3$.  This is because dark matter haloes collapse earlier and
merge together more rapidly in the overdense regions of the Universe that are
destined to form a rich cluster.  Although metals may be ejected, they are
quickly re--incorporated as the next level of the hierarchy collapses.  Note
that the metals in the cold gas also fall sharply to zero at low redshifts.
This is because galaxies that are accreted onto the cluster lose their supply
of new gas.  Stars continue to form and the cluster galaxies simply run out of
cold gas.

Comparison of Fig.~\ref{fig:budget} with Fig.~\ref{fig:budgetcl} suggests that
cosmic variance effects will turn out to be important when trying to constrain
different feedback schemes using estimates of the metallicity of the
intergalactic medium deduced from, for example, CIV absorption systems in the
spectra of quasars \citep{joop}. Nevertheless, the strong differences between
our different feedback schemes suggest that these kinds of measurements will
eventually tell us a great deal about how galaxies ejected their metals over
the history of the Universe.

\begin{figure*}
\bc
\resizebox{18cm}{!}{\includegraphics{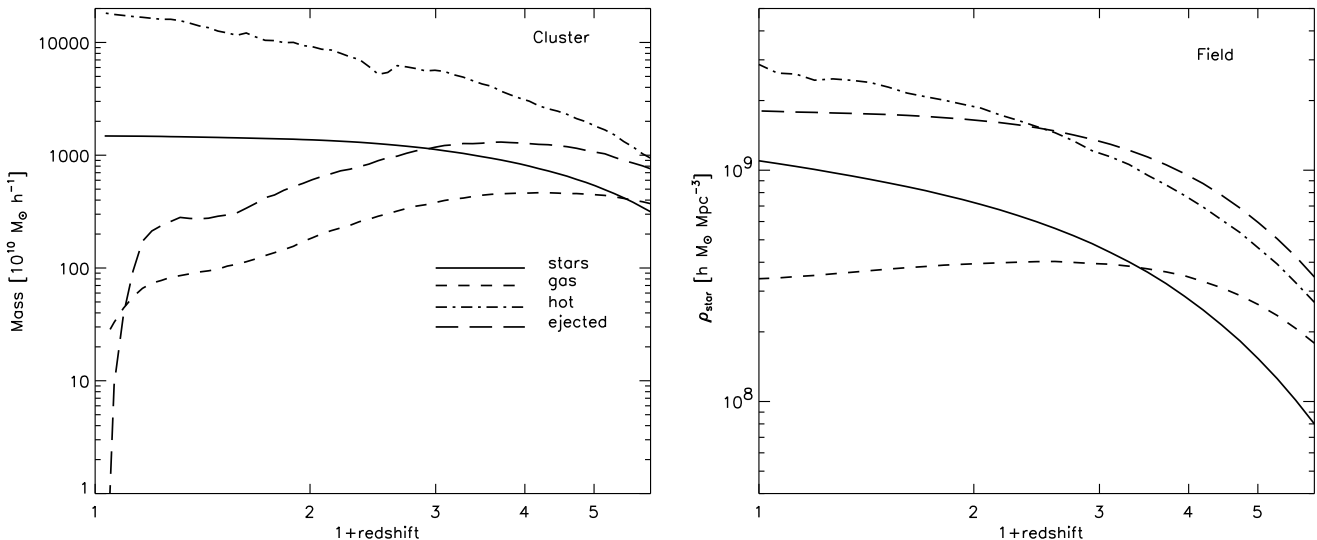}}\\%
\caption{Evolution of the different phases for the slow
  ejection scheme.  The left panel shows the results for the galaxies within
  the virial radius of our cluster simulation, while the right panel shows the
  results for a typical `field' region. }
\label{fig:budgetmass}
\ec
\end{figure*}

Finally, in Fig.~\ref{fig:budgetmass} we show the evolution of the different
phases for the slow ejection scheme, that can be considered as our `favourite'
model.  In the left panel we plot the evolution in mass of the different phases
for all the galaxies within the virial radius in our cluster simulation.  In
agreement with what is shown in Fig.~\ref{fig:sfr}, the stellar component grows
very slowly after redshift $\sim 3$, when most of the stars in the cluster have
already been formed.  The cold gas and the ejected mass decrease as galaxies
are accreted onto the cluster and stripped of their supply of new gas.  The hot
gas mass increases because of the accretion of `diffuse material'.  In the
right panel we plot the evolution of the mass density in each reservoir for our
field simulation.  Note that here the evolution in the stellar mass density is
more rapid than in the field.


\section{Discussion and Conclusions}
\label{sec:conclusions}

We have presented a semi--analytic model that follows the formation, evolution
and chemical enrichment of galaxies in a hierarchical merger model.  Galaxies
in our model are not closed boxes. They eject metals and we track the exchange
of metals between the stars, the cold galactic gas, the hot halo gas and an
ejected component, which we identify as the diffuse intergalactic medium (IGM).

We have explored three different schemes for implementing feedback processes in
our models:
\begin{itemize}
\item in the retention model, we assume that material reheated by supernovae
  explosions is ejected into the hot halo gas.
\item In the ejection model, we assume that this material is ejected outside
  the halo. It is later re--incorporated after a time that is of order the
  dynamical time of the halo.
\item In the wind model, we assume that galaxies eject material until they
  reside in haloes with $V_{\rm vir} > V_{\rm crit}$.  Ejected material is also
  re--incorporated on the dynamical time--scale of the halo.  The amount of gas
  that is ejected is proportional to the mass of stars formed.
\end{itemize}

In all cases we can adjust the parameters to obtain reasonably good agreement
between our model predictions and observational results at low redshift. The
wind scheme is perhaps the most successful, allowing us to obtain remarkably
good agreement with both the cluster luminosity function and the slope and
zero--point of the Tully--Fisher relation.  All our models reproduce a
metallicity mass relation that is in striking agreement with the latest
observational results from the SDSS.  By construction, we also reproduce the
observed trend of increasing gas fraction for smaller galaxies.  The good
agreement between models and the observational results suggests that we are
doing a reasonable job of tracking the circulation of metals between the
different baryonic components of the cluster.

We find that the chemical enrichment of the ICM occurs at high redshift:
$60$--$80$ per cent of the metals are ejected into the ICM at redshifts larger
than $1$, $35$--$60$ per cent at redshifts larger than $2$ and $20$--$45$ per
cent at redshifts larger than $3$.  About half of the metals are ejected by
galaxies with baryonic masses less than $1\times10^{10}\,h^{-1}{\rm
  M}_{\odot}$.  The predicted distribution of the masses of the ejecting
galaxies is very similar for all $3$ feedback schemes.  Although small galaxies
dominate the luminosity function in terms of number, they do not represent the
dominant contribution to the total stellar mass in the cluster.  Approximately
the same contribution to the chemical pollution of the ICM is from galaxies
with baryonic masses larger than $1\times10^{10}\,h^{-1}{\rm M}_{\odot}$.

Finally, we show that although most observations at redshift zero do not
strongly distinguish between our different feedback schemes, the observed
dependence of the baryon fraction on halo virial mass does place strong
constraints on exactly how galaxies ejected their metals.  Our results suggest
that gas and its associated metals must be ejected very efficiently from
galaxies and their associated dark matter haloes.  Once the material leaves the
halo, it must remain in the diffuse intergalactic medium for a time that is
comparable to the age of the Universe.

Future studies of the evolution of the metallicity of the intergalactic medium
should also be able to clarify the mechanisms by which such wind material is
mixed into the environment of galaxies.

\section*{Acknowledgements}
The simulations presented in this paper were carried out on the T3E
supercomputer at the Computing Center of the Max-Planck-Society in
Garching, Germany and on the IBM SP2 at CINES in Montpellier, France.  \\
We thank Jarle Brinchmann, Christy Tremonti, Tim Heckman, Sofia Alejandra Cora,
Volker Springel and Serena Bertone for useful and stimulating discussions.  We
thank Roberto De Propris for providing us with his data in electronic format
and Christy Tremonti for providing us with her data before publication.
Barbara Lanzoni, Felix Stoehr, Bepi Tormen and Naoki Yoshida are warmly thanked
for all the effort put in the re--simulation project and for letting us use
their simulations.

G.~D.~L. thanks the Alexander von Humboldt Foundation, the Federal Ministry of
Education and Research, and the Programme for Investment in the Future (ZIP) of
the German Government for financial support.

\bsp

\label{lastpage}

\bibliographystyle{mn2e}
\bibliography{ICM_delucia}

\renewcommand{\includegraphics}[1]{\ }

\end{document}